# Incoherent phonon transport dominates heat conduction across van der Waals superlattices

*Lu Zhao*[†,1], *Lijuan Zhang*[†,*,1], *Houfu Song*[1], *Hongda Du*[2], *Renshaw X. Wang*[3], *Junqiao Wu*[4,5], *Feiyu Kang*[*,1,2] *and Bo Sun*[*,1,2]

[1] Tsinghua-Berkeley Shenzhen Institute, Tsinghua University, Shenzhen, Guangdong 518055, China

[2] Institute of Materials Research, Tsinghua Shenzhen International Graduate School, Guangdong Provincial Key Laboratory of Thermal Management Engineering and Materials, Shenzhen, Guangdong 518055, China

[3] School of Physical and Mathematical Sciences, Nanyang Technological University, Singapore 637371, Singapore

[4] Department of Materials Science and Engineering, University of California, Berkeley, Berkeley, CA 94720, USA

[5] Materials Sciences, Lawrence Berkeley National Laboratory, Berkeley, CA 94720, USA

[†] These authors contributed equally to this work

*Email: zhang.lijuan@sz.tsinghua.edu.cn, fykang@sz.tsinghua.edu.cn, sun.bo@sz.tsinghua.edu.cn

Abstract. Heat conduction mechanisms in superlattices could be different across different types of interfaces. Van der Waals superlattices are structures physically assembled through weak van der Waals interactions by design, and may host properties beyond the traditional limits of lattice matching and processing compatibility, offering new types of interfaces. In this work, natural van der Waals $(SnS)_{1.17}(NbS_2)_n$ superlattices are synthesized, and their thermal conductivities are measured by time-domain thermoreflectance as a function of interface density. Our results show that heat conduction of $(SnS)_{1.17}(NbS_2)_n$ superlattices is dominated by interface scattering when the coherent length of phonons is larger than the superlattice period, indicating incoherent phonon transport dominates cross-plane heat conduction in van der Waals superlattices even when the period is atomically thin and abrupt. Moreover, our result suggests that the widely accepted heat conduction mechanism for conventional superlattices that coherent phonons dominate when the period is short, is not applicable due to symmetry breaking in most van der Waals superlattices. Our findings provide new insight for understanding the thermal behavior of van der Waals superlattices, and devise approaches for effective thermal management of superlattices depending on the distinct types of interfaces.

Van der Waals (vdW) superlattices and heterostructures, vertically assembled using different monolayers or few-layer two-dimensional (2D) crystals, are emerging as a promising material system for electronic, photonic and plasmonic studies.[1-6] Similar to conventional superlattices, interfaces play a key role in the behavior of vdW superlattices.[7,8] However, unlike conventional superlattices that are strongly covalently bonded across the interface, vdW superlattices are physically stacked together through weak van der Waals interactions, which allow rational design and flexible choices of combining highly distinct layers without consideration of lattice matching and processing compatibility, and thus, represent a distinctly different type of superlattice and interface.[9]

Thermal transport across interfaces is of vital importance in a diverse range of applications due to the increasing importance of thermal management, especially in devices based on superlattices and heterostructures where interface thermal transport typically dominates the heat conduction. In the past several decades, thermal transport in conventional superlattices has been extensively studied.[10-18] For example, Koh and Cahill[11] reported that interface thermal conductance increases with reduced superlattice period, suggesting that long-wavelength phonons are the dominant heat carriers in short-period AlN/GaN superlattices. In addition, Ravichandran et al.[12] observed that there is a crossover from incoherent to coherent phonon transport in oxide superlattices grown by molecular-beam epitaxy (MBE) when the coherent length of phonons is comparable to the period of the lattice. Despite these advances, it is unknown whether these heat conduction mechanisms are applicable to vdW superlattices and heterostructures, because the interface in these structures is fundamentally different.

When studying thermal transport across the vdW superlattices, it is challenging to obtain the intrinsic phonon transport behavior across interfaces, because the bottom-up assembled vdW superlattices normally suffer from interface contamination and turbostratic disorders.[19-23] Contamination-free and well-ordered vdW superlattices are required to study intrinsic phonon transport properties across these vdW interfaces. The family of misfit layered compounds are naturally occurring van der Waals superlattices that meet such requirements. They have a generic formula of $(MX)_{1+\delta}(TX_2)_n$, where M = Sn, Pb, Sb, Bi, or a rare earth element, T = group IV or V transition metals, X = S or Se, and $\delta$ is the degree of structural mismatch between MX and $TX_2$.[24-

[26] In $(MX)_{1+\delta}(TX_2)_n$, each period is consisted of monolayer $(MX)_{1+\delta}$ and n layers of $TX_2$, where n can be tuned to adjust interface density. In addition, misfit layered compounds is a promising family of thermoelectric materials. It is thus worthwhile to study their thermal conduction properties.[27,28]

Here, we report that cross-plane heat conduction in van der Waals superlattices is dominated by incoherent phonon transport. Using misfit layered compounds $(SnS)_{1.17}(NbS_2)_n$, we construct vdW superlattices with atomically sharp, contamination-free interfaces, which enable us to study the intrinsic phonon transport mechanism across the vdW interfaces. We show that incoherent phonons dominate heat transport in this system, despite that the phonon coherent length is larger than the superlattice period. This finding is in stark contrast to heat conduction in conventional AlAs/GaAs, BTO/STO and naturally occurred perovskite superlattices, where coherent phonon dominates when the period is short.[12,13,17,18] Our work provides new insights for understanding of thermal behaviour of vdW superlattices and heterostructures, and devises approaches for effective thermal management across different types of interfaces.

## Results and Discussion

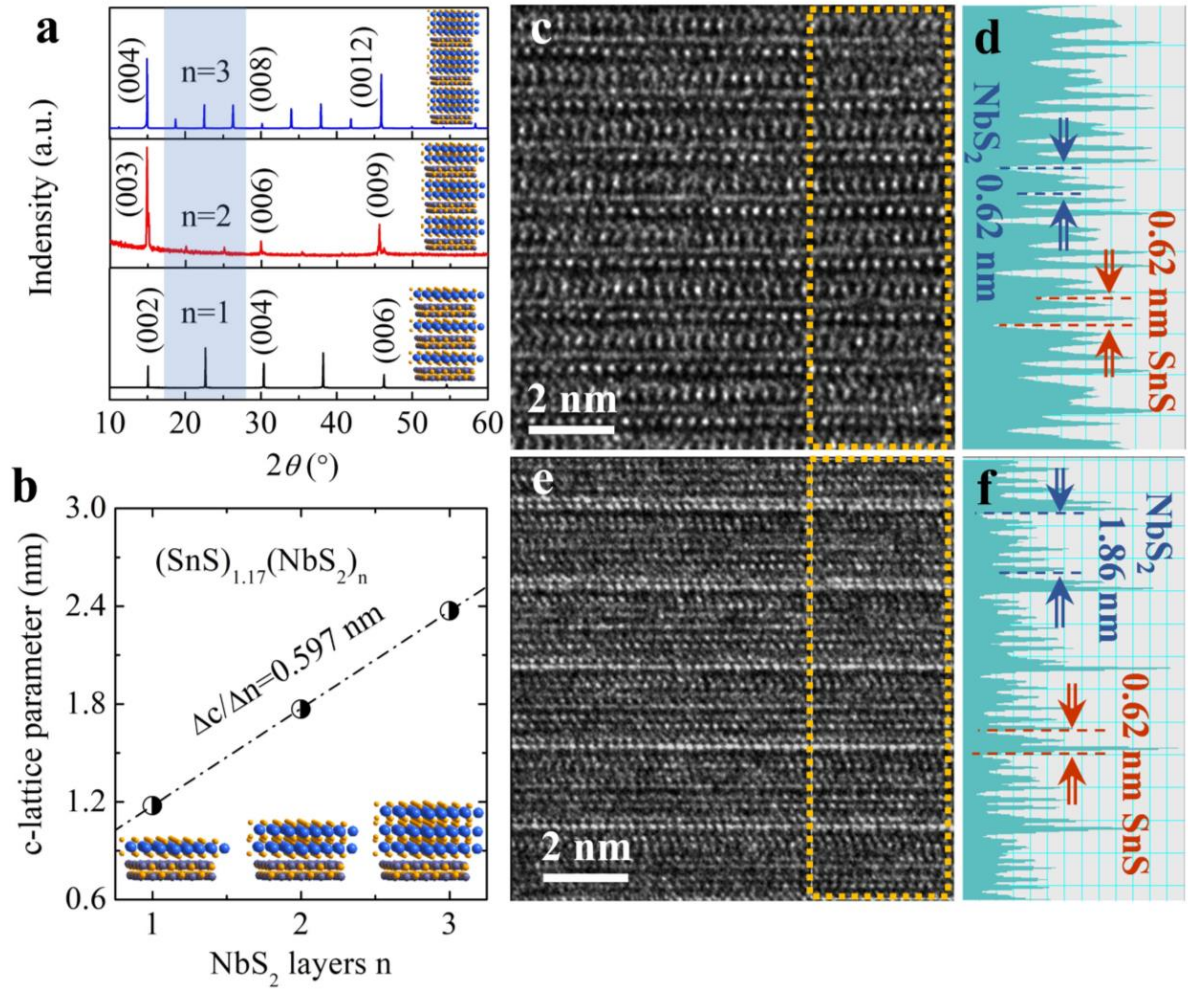

**Figure 1**. (a) X-ray diffraction patterns for the $(SnS)_{1.17}(NbS_2)_n$ series. Selected (00l) reflections are labelled in the figure. The shaded region indicates consecutive structural orders were obtained as the number of inserted NbS$_2$ layer increases. (b) c-lattice parameter as a function of the number of NbS$_2$ layers. Representative HRTEM images for (c-d) $(SnS)_{1.17}NbS_2$; (e-f) $(SnS)_{1.17}(NbS2)_3$ samples along the [010] direction. (d) and (f): line profile images for $(SnS)_{1.17}NbS_2$ and $(SnS)_{1.17}(NbS_2)_3$ superlattices highlighted by white rectangle in (c) and (e), respectively.

We select $(SnS)_{1.17}(NbS_2)_n$ (n = 1, 2, 3) misfit layered compounds for this study, which were synthesized by chemical vapour transport.[26] The structures of as-grown $(SnS)_{1.17}(NbS_2)_n$ (n = 1, 2, 3) are shown in Figure 1a,b. All of the Bragg peaks are sharp and can be indexed to the (00l) family of planes, suggesting high crystallinity and preferred crystallographic alignment of the layers. As the number of inserted $NbS_2$ layer increases, the diffraction peaks present higher degree of consecutive structural orders (highlighted by the light-blue rectangle in Figure 1a). The existence of well-ordered, consecutively increased diffraction peaks indicates superlattice structures for $(SnS)_{1.17}(NbS_2)_n$ were formed. Additional insertion of $NbS_2$ layer results in a systematic increase in c-axis lattice parameter by 0.597 nm (Figure 1b). This result is consistent with the expected repeat unit thickness of monolayer of $NbS_2$ (0.598 nm).[29] In the inset of Figure 1b we schematically illustrate the structure of $(SnS)_{1.17}(NbS2)_n$, which shows the incommensurate sublattices.

High-resolution transmission electron microscopy confirms that $(SnS)_{1.17}NbS_2$ superlattice has the expected layering structure, which consists of vertically stacked SnS and $NbS_2$ sublayers, with atomically sharp and contamination-free interfaces (Figure 1c,d). The natural vdW superlattice feature and spontaneous growth process lead to these high-quality interfaces.[30-33] Brighter regions correspond to greater scattering due to heavier atomic mass. The area of SnS, therefore, appears brighter than that of $NbS_2$. These two substructures display the same repeat thickness, 0.62 nm, according to the line profile (Figure 1d). The interfaces could provide considerable barriers to scatter heat-carrying phonons, despite that the coherent length of phonons is larger than the period of vdW superlattices, which will be discussed later. $(SnS)_{1.17}(NbS_2)_3$ superlattice consists of one-by-three vertically stacked SnS (0.62 nm) and $NbS_2$ (1.86 nm) sublayers as shown in Figure 1e,f. The additional increase in repeat unit thickness (0.62 nm) is slightly larger than the result (0.597 nm) obtained from x-ray diffraction (Figure 1b). No obvious turbostratic disorder was observed in the $(SnS)_{1.17}(NbS_2)_n$ series.[23]

We measured the cross-plane thermal conductivity, $\Lambda$, of the $(SnS)_{1.17}(NbS_2)_n$ series as a function of temperature using time-domain thermoreflectance (TDTR),[34,35] which has been widely used in measuring thermal conductivity of superlattices, 2D materials etc.[36-39] Details of the TDTR measurements are presented in the Methods section in Supporting Information. In Figure 2a, $\Lambda$ reduces with decreasing temperature for all $(SnS)_{1.17}(NbS_2)_n$ samples, which is typical for long-period superlattices, indicating that the dominate phonon scattering mechanism is boundary or interface scattering.[11,12] Moreover, we find that $\Lambda$ reduces monotonically as the interface density increases. At room temperature, $\Lambda$ for bulk single crystalline SnS and $NbS_2$ are 1.37 and 1.64 W $m^{-1}$ $K^{-1}$, and decreases to 0.82, 0.62 and 0.44 W $m^{-1}$ $K^{-1}$ for $(SnS)_{1.17}(NbS_2)_3$, $(SnS)_{1.17}(NbS_2)_2$ and $(SnS)_{1.17}NbS_2$ as the interface density increases to 0.81 (n = 3), 1.08 (n = 2), and 1.61 $nm^{-1}$ (n = 1), respectively. The calculated coherence length of SnS for longitudinal and transverse modes is 6.2 and 5 nm, respectively (Details of calculations are available in the Methods section, Supporting Information). As the interface density increases, the cross-plane thermal conductivity at room temperature continuously reduces, indicating that the heat conduction is dominantly limited by interface scattering, even when the superlattice period is atomically thin and shorter than the coherent length of phonons (Figure 2b). This result is in contrary to the mechanism of heat conduction observed in AlAs/GaAs and BTO/STO superlattices, where coherent phonons dominate the heat conduction when the superlattice period is short.[12,13]

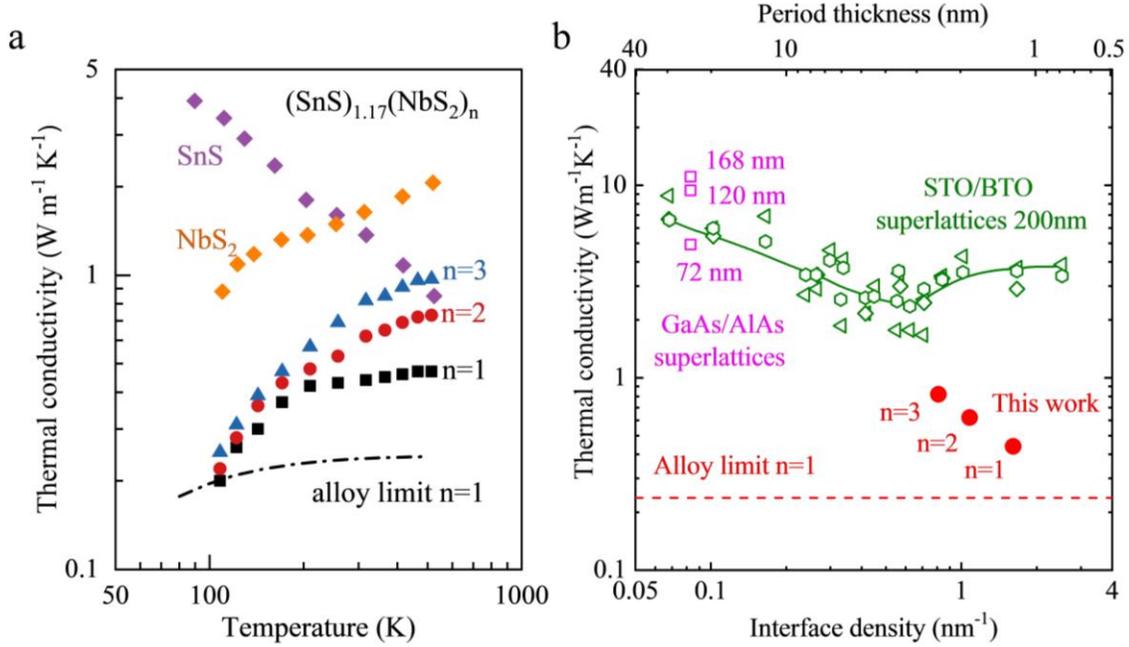

**Figure 2.** (a) Cross-plane thermal conductivity as a function of temperature for the $(SnS)_{1.17}(NbS_2)_n$ series. The black dashed line refers to the alloy limit of $(SnS)_{1.17}NbS_2$ calculated using the Cahill-Pohl model.[40] The purple and orange diamond refers to the bulk SnS and $NbS_2$, and $T^{-1}$ dependence of SnS suggests that phonon-phonon scattering dominate the heat transport. (b) Measured room-temperature thermal conductivity values for various superlattices as a function of interface density. The room-temperature thermal conductivity of GaAs/AlAs superlattice[13] and $(SrTiO_3)_m/(BaTiO_3)_n$ superlattice[12] are included for comparison.

In order to obtain more insights about the heat conduction mechanism in the vdW superlattices, the thermal conductance G across the $SnS/NbS_2$ interface is estimated using the following formula:[11]

$$\frac{2}{G} = \frac{h_{SnS} + h_{NbS_2}}{\Lambda_{superlattice}} - \frac{h_{SnS}}{\Lambda_{SnS_2}} - \frac{h_{NbS_2}}{\Lambda_{NbS_2}} \qquad (1)$$

where $h_{SnS}$, $h_{NbS_2}$, $\Lambda_{superlattice}$, $\Lambda_{SnS_2}$, and $\Lambda_{NbS_2}$ are the thickness of the SnS sublayer, thickness of the NbS$_2$ sublayer, thermal conductivity of the entire superlattice, of SnS[41] and of NbS$_2$, respectively. The derived room-temperature G are plotted in Figure 3. As the interface density increases, the thermal conductance G shows a monotonous reduction. At room temperature, the thermal conductance G for the sample with interface density of 0.81 nm$^{-1}$ (n = 3) is 4.21 GW m$^{-2}$ K$^{-1}$, and it decreases to 1.33 GW m$^{-2}$ K$^{-1}$ as the interface density increases to 1.61 nm$^{-1}$ (n = 1). This could be explained as long-wavelength phonons being scattered more strongly when the interface density is large, which is in opposite to the coherent phonon scattering behavior in conventional superlattices.[11,42-44] The considerable decrease in interface thermal conductance indicates that the major scattering mechanism is interface scattering in (SnS)$_{1.17}$(NbS$_2$)$_n$ superlattices (Figure 3).

One of the important predictions regarding interface thermal conductance G in short-period superlattices is that G increases with the reduction in superlattice period due to the coherent transport of long-wavelength phonons.[11] Such behavior has been observed in short-period Si/Si$_{0.7}$Ge$_{0.3}$ superlattices,[44] GaAs/AlAs superlattices,[42] W/Al$_2$O$_3$ multilayers,[43] and perovskite superlattices.[17] However, in our misfit vdW superlattices, we see the opposite trend that G reduces following the reduction of superlattice period, even though the interface is atomically thin and smaller than the coherent length of phonons. We postulate that this obvious contrast stems from the difference in bonding types at the interfaces. Conventional fabrication for superlattices and heterostructures by epitaxial growth relies on strong covalent bonding across the interface. Symmetry, lattice mismatching and deposition induced strain/defects at and beyond the interfaces result in a strong scattering of phonons in the propagation of phonons across epitaxy superlattices.[45,46] As the coherent length of phonons is comparable or larger than the superlattice period, long wavelength phonons will transport coherently across the interface in conventional superlattices, resulting in enhancement of thermal transport.[11,42-44] However, in our vdW superlattices, the interface is weakly bonded by the van der Waals interactions, and the lattice symmetry is broken across the incommensurate sublayers, although the superlattice period is atomically thin. Thus, phonons will lose their coherence when propagating through such interfaces, leading to the reduction of thermal conductance and an incoherent phonon transport.

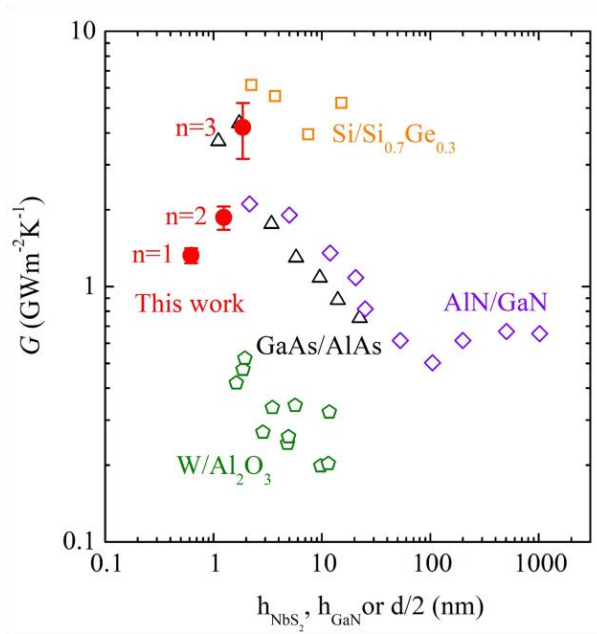

**Figure 3**. Room-temperature interface thermal conductance of $(SnS)_{1.17}(NbS_2)_n$ series (solid circles). The interface thermal conductance of AlN/GaN superlattices (open diamond)[11], $Si/Si_{0.7}Ge_{0.3}$ superlattices (open square), GaAs/AlAs superlattices (open triangles)[42], and $W/Al_2O_3$ multilayers (open pentagon )[43] are included for comparison and are plotted as a function of $h_{GaN}$ or half of the period, d/2.

Our result is fundamentally different from the artificial $MoS_2/WS_2$ vdW superlattices calculated by Guo and Minnich[47], where they computed that phonon can transport coherently across interfaces. In the case of $MoS_2/WS_2$, the lattice constant of $MoS_2$ (one layer: a = 0.3125 nm; two layers: a = 0.3126 nm) and $WS_2$ (one layer: a = 0.3125 nm; two layers: a = 0.3125 nm) are nearly the same so they can form superlattices with minimum lattice mismatch.[48] Thus, the interfaces in $MoS_2/WS_2$ superlattice maintain lattice symmetry that could host coherent phonon transport. In our case, the lattice mismatch between SnS (a = 0.5673 nm) and $NbS_2$ (a = 0.3321 nm)[24] is large, and hence, phonon transport in $(SnS)_{1.17}(NbS_2)_n$ superlattices is governed by the discontinuity in symmetry, which destroys the wave interference and changes the nature of phonon transport from coherent to diffusive when the phonons propagate across the interfaces.[49] Our results indicate that incoherent phonons dominate cross-plane heat conduction, even when the period is atomically thin, clean and abrupt, for lattice mismatched vdW superlattices. In real-world applications, vdW superlattices and heterostructures are predominately lattice mismatched,[50] and thus our results and conclusion could be applied to the broad family of vdW superlattices and heterostructures.

**Conclusions**

In summary, natural van der Waals $(SnS)_{1.17}(NbS_2)_n$ superlattices are synthesized by chemical vapour transport. The cross-plane thermal conductivity, $\Lambda$, of the $(SnS)_{1.17}(NbS_2)_n$ series are measured by time-domain thermoreflectance, showing a continuous reduction as the interface density increases. The continuous reduction of thermal conductivity is related to additional interfacial thermal resistance induced by the insertion of more $NbS_2$ layers, which reveals that coherent phonon transport is not important in heat conduction in vdW superlattices. In contrast to conventional covalent superlattices where coherent phonon dominates when the period is short, our result of incoherent phonon transport dominating heat conduction in vdW superlattices provides a new and broadly applicable insight that heat transport mechanisms in superlattices and heterostructures are different depending on the specific physical property of interfaces.

**Methods**

**Sample Preparation.** Van der Waals $(SnS)_{1.17}(NbS_2)_n$ superlattices were synthesized by iodine vapour transfer by mixing appropriate ratios of high-purity Sn, Nb, S and iodine in vacuum quartz tubes. Single crystalline SnS and $NbS_2$ were purchased from HQ graphene. The size of these flakes were at least 3mm× 3mm. Samples were kept in the argon filled glove box before Aluminum deposition and measurement to minimize the possibility of oxidation and contamination.

**TDTR Measurement.** The temperature dependent thermal conductivity of superlattices were measured in Janis ST-500 vacuum cryostat by ultrafast laser-based time-domain thermoreflectance (TDTR). All samples were coated with 80-nm-thick Aluminum film by e-beam evaporation to act as the transducer, and the surface reflectance changed linearly with temperature under periodic pump beam heating, and the probe beam detected the corresponding response as a function of delay time between the pump and probe pulses.

**Alloy limit and coherence length calculations.** The alloy limit was calculated by Cahill-Pohl model and considered as the minimum phonon thermal conductivity for fully dense amorphous and disordered materials, in which model, the phonons were assumed as the localized quantum oscillators and random walking without phase coherence. The coherence length calculated by a method reported elsewhere (see Supporting Information) was used to describe the characteristic scale for distinguishing phonon coherent transport and diffusive transport.

**Supporting Information.**

Supporting Information is available from the ACS Publications or from the author.

**Corresponding Author**

*Email: zhang.lijuan@sz.tsinghua.edu.cn, fykang@sz.tsinghua.edu.cn,

sun.bo@sz.tsinghua.edu.cn

**Author Contributions**

B.S. conceived and supervised the project with J.W. and F.K. L. Zhang grew the samples with help from H.D. and R.X.W., L.Zhao performed the TDTR measurements with H.S. B.S., L.Zhang and L.Zhao wrote the manuscript with input from all authors. †These authors contributed equally.

**Acknowledgements**

B.S., L.Zhang, L.Zhao and H.S. acknowledge the support of Guangdong Natural Science Foundation (No. 2019A1515010868), the National Natural Science Foundation of China (No. 12004211), the Shenzhen Science and Technology Program (No. RCYX20200714114643187) and Tsinghua Shenzhen International Graduate School (No. QD2021008N). L.Zhang acknowledges the the support from the National Natural Science Foundation of China (No. 52002206), the Postdoctoral Science Foundation (No. 2019M650669) and the Shenzhen Science and Technology Program (No. RCBS20200714114857131). The authors thank the support from the Testing Technology Center of Materials and Devices in Tsinghua Shenzhen International Graduate School and Analytical Instrumentation Center in Peking University Shenzhen Graduate School.
13

**Table of contents**

Interface dominates heat conduction in vdW superlattices even when the phonon coherent length is larger than superlattice period. In contrast to the conventional, covalent superlattices where coherent phonon dominates when the period is short, our findings provide new insight for understanding the thermal behavior in vdW heterostructures, and devise approaches for effective thermal management across different types of interfaces.

Lu Zhao[†], Lijuan Zhang[†,*], Houfu Song, Hongda Du, Renshaw X. Wang, Junqiao Wu, Feiyu Kang* and Bo Sun*

**Incoherent phonon transport dominates heat conduction across van der Waals superlattices**

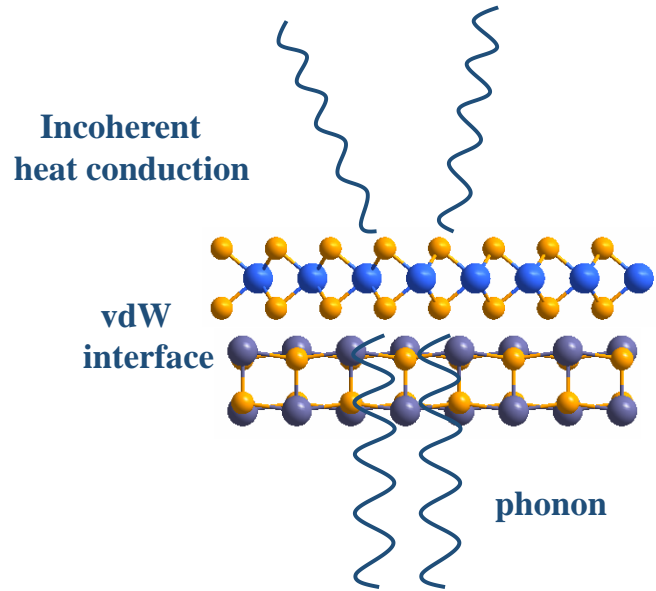

Supporting Information

**Incoherent phonon transport dominates heat conduction across van der Waals superlattices**

*Lu Zhao*[†,1], *Lijuan Zhang*[†,*,1], *Houfu Song*[1], *Hongda Du*[2], *Renshaw X. Wang*[3], *Junqiao Wu*[4,5], *Feiyu Kang*[*,1,2] *and Bo Sun*[*,1,2]

[1] Tsinghua-Berkeley Shenzhen Institute, Tsinghua University, Shenzhen, Guangdong 518055, China

[2] Institute of Materials Research, Tsinghua Shenzhen International Graduate School, Guangdong Provincial Key Laboratory of Thermal Management Engineering and Materials Shenzhen, Guangdong 518055, China

[3] School of Physical and Mathematical Sciences, Nanyang Technological University, Singapore 637371, Singapore

[4] Department of Materials Science and Engineering, University of California, Berkeley, Berkeley, CA 94720, USA

[5] Materials Sciences, Lawrence Berkeley National Laboratory, Berkeley, CA 94720, USA

[†] These authors contributed equally to this work

*Email: zhang.lijuan@sz.tsinghua.edu.cn, fykang@sz.tsinghua.edu.cn, sun.bo@sz.tsinghua.edu.cn

1. **Sample preparation and characterization**

**Synthesis.** Flakes of $(SnS)_{1.17}(NbS_2)_n$ (n = 1, 2, 3) were synthesized by iodine vapour transfer. Stoichiometric elementary substances were sealed in vacuum quartz by mixing appropriate ratios of high-purity Sn (Aladdin, 99.99%), Nb (Macklin, 99.99% ), S (Aladdin, 99.99%) and iodine (Macklin, 99.99%) in 17-mm-diameter quartz tubes in an argon filled glove box. The tubes were then evacuated and flame-sealed. Sealed quartz with starting materials of $(SnS)_{1.17}NbS_2$ was kept at a temperature gradient between 850 and 770 °C for four weeks. $(SnS)_{1.17}(NbS_2)_2$ and $(SnS)_{1.17}(NbS_2)_3$ flakes were sintered at a temperature gradient between 950 and 900 °C for four weeks. In a typical experiment, 0.1667 g of Sn, 0.1464 g of Se, 0.1115 g of Nb and 0.0191 g of iodine were used to prepare $(SnS)_{1.17}NbS_2$. Plate like crystals up to 3 mm length were obtained. And single crystalline SnS and $NbS_2$ were purchased from HQ graphene.

**X-ray diffraction (XRD).** The XRD patterns were obtained with Cu Kα ($\lambda$ = 1.5418 Å) radiation by Bruker D8 Advance and equipped with LynxEye detector operating at 40 kV and 40 mA. The studies were performed in conventional Bragg-Brentano configuration for the range of 2θ from 10° to 60° with step size of 0.02°.

**High-resolution transmission electron microscopy (HRTEM).** TEM was performed using a JEOL JEM 3200FS microscope with an accelerating voltage of 300 kV. The thin TEM specimens were fabricated by focused ion beam (FIB) on FEI Scios at a moderate voltage (30 keV), which was then followed by cleaning with a lower voltage (5 keV and 2 keV).

2. **Thermal property measurement**

**Principles of TDTR.** Thermal conductivity in the cross-plane direction of the superlattices was measured by time-domain thermoreflectance (TDTR) for SnS and NbS$_2$ bulk and (SnS)$_{1.17}$(NbS$_2$)$_n$ (n = 1, 2, 3) series. In TDTR, a pump laser is directed to periodically heat up the surface of a sample, and a temporally delayed probe laser is used to track the temperature evolution of the surface (by detecting the reflectance change) as the surface temperature cools down. We use the ratio (R) of the in-phase ($V_{in}$) and out-of-phase ($V_{out}$) components of the thermoreflectance signal detected by the lock-in amplifier to monitor the surface temperature decay over the full time delay range.[1-4] The ratio R=-$V_{in}$/$V_{out}$ is used to extract the thermal conductivity in conditional TDTR. All the samples were coated with a thin Aluminum film that acts as temperature transducer by thermal evaporation. We measure the thickness of the Al films in each measurement with picosecond ultrasonics. In this study, the modulation frequency of the pump beam was fixed at 10.1 MHz. The radii of the pump and probe beam are 10.3 μm at the sample surface. We use total laser powers of 6-10 mW, creating temperature rises of <10 K.

**Sensitivity.** The sensitivity coefficient $S_\alpha$ represents the extent to which each parameter affects the measured signal R. Thus $S_\alpha$ can be quantitatively analysed through the following equation.

$$S_\alpha = \frac{\partial \ln R}{\partial \ln \alpha} = \frac{\alpha}{R} \frac{\partial R}{\partial \alpha}$$

Where α is the parameter used in the thermal model to fit the measured signal R, such as the thickness of Al, the thermal conductivity and heat capacity of superlattice. Sensitivity plot is also a powerful tool to analysis the design of experiments to achieve the most accurate results. In this study, the modulation frequency of the pump beam was fixed at 10.1 MHz. The radii of the pump and probe beam are 10.3 μm at the sample surface. So that the thermal penetration depth $d$ ($d = \sqrt{\Lambda/\pi C f}$, where $\Lambda$ is the thermal conductivity, C is the volumetric heat capacity and $f$ is the modulation frequency) is much smaller than the thickness of supperlattice. We use total laser powers of 6-10 mW, creating temperature rises of <10 K.

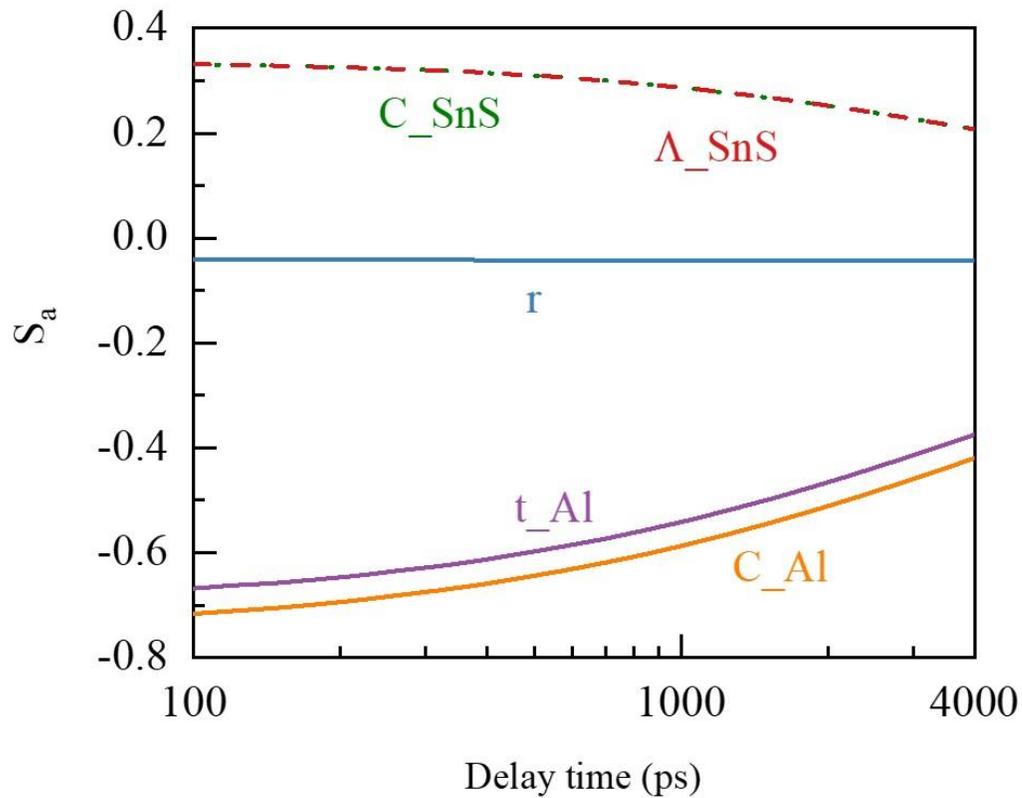

**Figure S1**. Sensitivity plot of volumetric heat capacity of SnS (green dashed line), thermal conductivity of SnS (red dashed line), spotsize of laser beam (blue line), thickness of Al film (purple line) and volumetric heat capacity of Al (orange line) in TDTR measurement.

**Thermal model and fitting**. Thermal conductivity can be derived by fitting the measured signal R with the multilayer thermal model, which has been explained in detail in SI Reference 2.

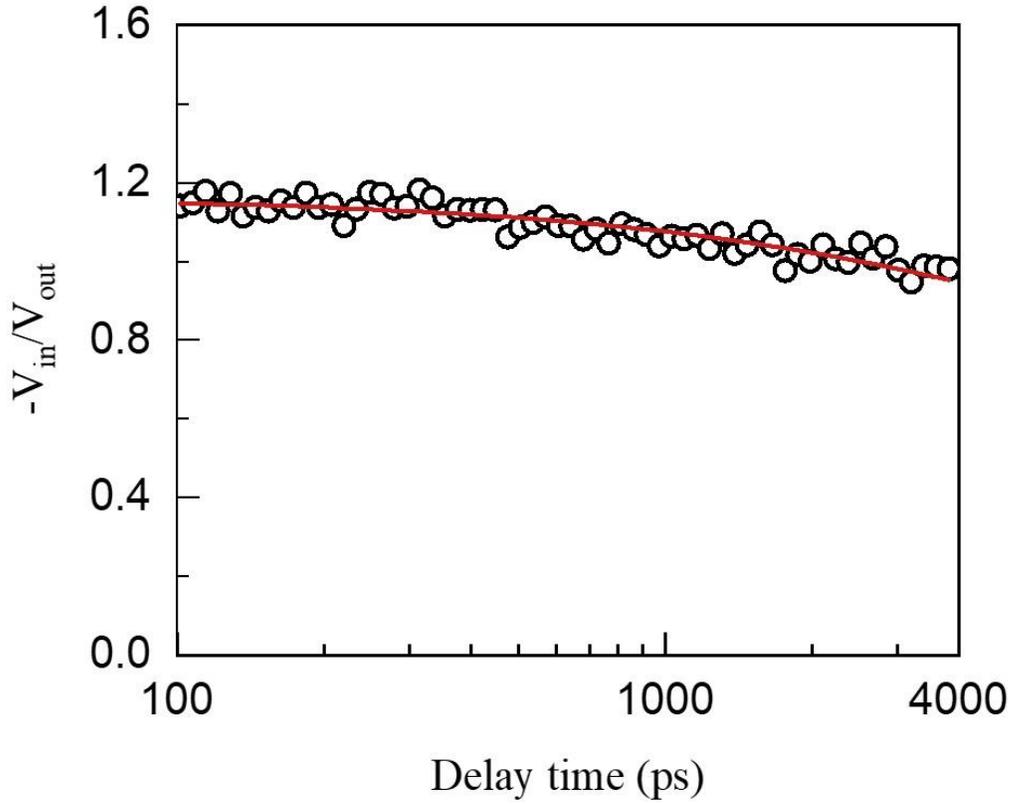

**Figure S2**. Best fitting for measured signal of SnS.

**Uncertainty.** After sensitivity calculations for all measurement parameters, the uncertainty can be evaluated by the following equation.

$$\left(\frac{\Delta \Lambda}{\Lambda}\right)^2 = \sum \left(\frac{S_\alpha}{S_\Lambda}\frac{\Delta \alpha}{\alpha}\right)^2 + \left(\frac{S_\phi}{S_\Lambda}\delta_\phi\right)$$

Where $\frac{\Delta \alpha}{\alpha}$ is the uncertainty of the measurement parameter α, and $\delta_\phi$ is the phase uncertainty. The total uncertainty of TDTR measurement for bulk SnS, $NbS_2$ and superlattice is lower than 7%.

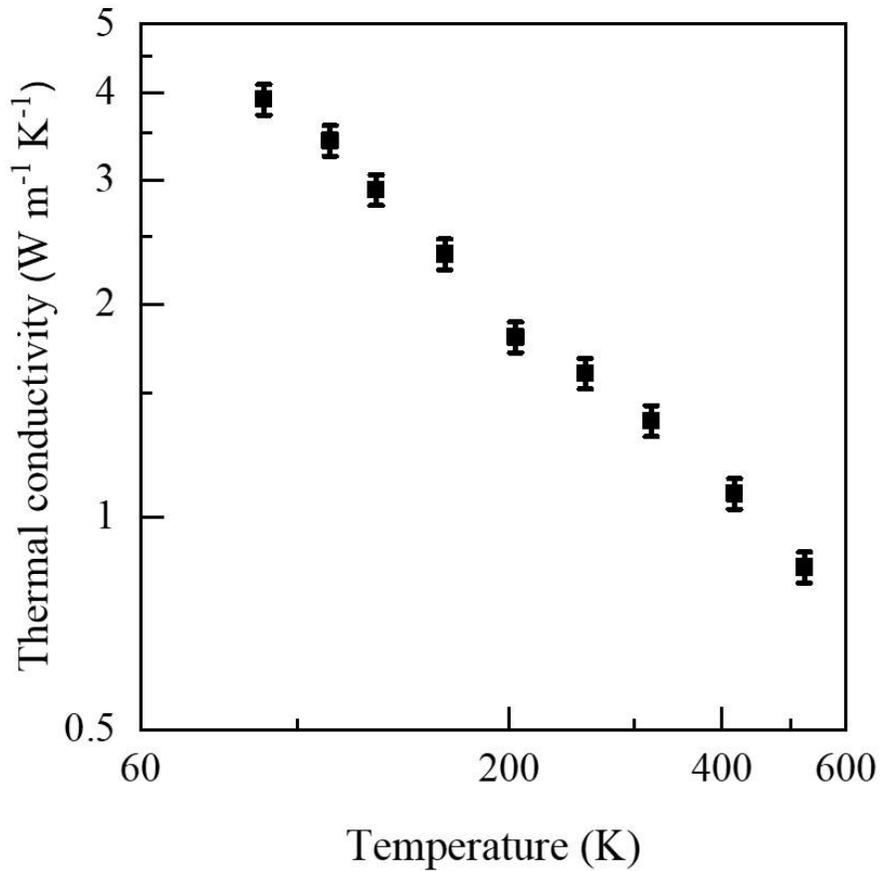

**Figure S3.** Temperature dependent cross-plane thermal conductivity with error bar of SnS.

**Coherence length calculations.** To compare with the conventional covalent superlattices where coherent phonon dominates when the period is short,[5] we should calculate the phonon coherence length of $(SnS)_{1.17}(NbS_2)_n$ superlattices to predict whether the coherent heat transport occurs at or beyond their phonon coherent length. The phonon coherence length calculation was performed using a method reported elsewhere.[6] An averaged Debye-like dispersion was assumed for all the acoustic branches; the contributions of the optical branches were ignored.[7] Here, we use the coherence length of SnS as a qualitative estimate,[7] and it is important to realize that one must consider the full dispersion for fully accurate coherence-length calculations. The zone edge cut-off frequencies and group velocities for SnS were obtained from appropriate references.[8]